# Titanium dioxide hole-blocking layer in ultra-thin-film crystalline silicon solar cells

Y. Kang[1,4,*], H. Deng[1,4,*], Y. Chen[1], Y. Huo[1], J. Jia[1], L. Zhao[2], Z. Zaidi[1], K. Zang[1], J. S. Harris[1,2,3]

1. Department of Electrical Engineering, Stanford University, Stanford, California 94305, USA
2. Department of Material Science and Engineering, Stanford University, Stanford, California 94305, USA
3. Department of Applied Physics, Stanford University, Stanford, California 94305, USA
4. These two authors contribute equally in this work.
* Authors to whom correspondence should be addressed.
   Email: kys86@stanford.edu and huiyang.deng@stanford.edu

**One of the remaining obstacles to approaching the theoretical efficiency limit of crystalline silicon (c-Si) solar cells is the exceedingly high interface recombination loss for minority carriers at the Ohmic contacts. In ultra-thin-film c-Si solar cells, this contact recombination loss is far more severe than for traditional thick cells due to the smaller volume and higher minority carrier concentration of the former. This paper presents a novel design of an electron passing (Ohmic) contact to n-type Si that is hole-blocking with significantly reduced hole recombination. This contact is formed by depositing a thin titanium dioxide ($TiO_2$) layer to form a silicon metal-insulator-semiconductor (MIS) contact. A 2 μm thick Si cell with this $TiO_2$ MIS contact achieved an open circuit voltage (Voc) of 645 mV, which is 10 mV higher than that of an ultra-thin cell with a metal contact. This MIS contact demonstrates a new path for ultra-thin-film c-Si solar cells to achieve high efficiencies as high as traditional thick cells, and enables the fabrication of high-efficiency c-Si solar cells at a lower cost.**

Ultra-thin-film crystalline silicon (c-Si) solar cells have recently attracted great interest due to their potential to lower the cost and increase the efficiency of c-Si solar cells. However, current state-of-the-art ultra-thin-film c-Si solar cells have performed well below the projected efficiency [1-4]. As the thickness of solar cells decreases, the recombination loss in the bulk Si is substantially reduced, and the recombination loss at surfaces, especially the metal Ohmic contacts, becomes the main obstacle for high performance. The key to solve this issue is forming carrier-selective contacts in c-Si solar cells, which can provide both lower minority carrier recombination velocity and more efficient majority carrier transport. Traditional c-Si solar cells use a diffused emitter and back surface field (BSF) to create carrier-selective layers. However, it has been shown that recombination at the metal/silicon interface still causes more than 40% of the total recombination losses [5,6]. To address the challenge of further controlling the contact recombination loss, various advanced cell designs have been demonstrated. One example is the heterojunction solar cell with an intrinsic larger bandgap thin-film (HIT) [7-9]. However, the a-Si:H layer in HIT cells has high parasitic absorption and high defect concentrations that results in a loss in the short circuit current ($J_{sc}$) [10,11]. Another method is to deposit a thin tunneling silicon dioxide ($SiO_2$) layer as a carrier selective contact [12-15]. However, to achieve low contact resistance, the thickness of the tunneling $SiO_2$ layer has to be very precisely controlled, and challenging for large-scale manufacturing. Recently, c-Si solar cells with Si/organic and/or Si/metal-oxide heterojunctions [16,17] have been demonstrated, providing an alternative solution of carrier-selective layer. However, those designs suffer the quality of surface passivation and the low $V_{oc}$.

In this paper, we demonstrate a carrier-selective contact formed by depositing a thin layer of $TiO_2$ between the metal and n-type Si to form a metal-insulator-semiconductor (MIS) contact, which can effectivity block holes without compromising the conductivity for electrons. $TiO_2$ has a large valance band (VB) offset ($\Delta E_v$), resulting in holes being blocked. Meanwhile, good passivation can be provided by $TiO_2$ on an n-type Si surface, achieving a surface recombination velocity (SRV) as low as 100 cm-sec$^{-1}$ [18-20]. TiO$_2$ can thus prevent holes from recombining at this interface or diffusing into the metal. For the conduction band (CB), TiO$_2$ can unpin the Fermi level at the Si surface, eliminate the Schottky barrier, and pin the Fermi level of the metal very close to the CB edge of Si [21,22]. This MIS contact thus has a small CB offset ($\Delta E_c$), allowing electrons to transport freely through the hole blocking layer. With the aligned CBs, this MIS contact also has good tolerance for thickness variations of the TiO$_2$ layer. We experimentally demonstrate TiO$_2$ carrier-selective contact on an ultra-thin-film c-Si junction solar cell, which can reduce the SRV at the contacts and suppress the recombination current. A 2 μm thick c-Si solar cell with the TiO$_2$ MIS hole-blocking contact achieves a $V_{oc}$ of 645mV, which is significantly higher than recently published c-Si cells of similar thickness [1-4].

To demonstrate the benefits of carrier-selective contacts in solar cells, we simulated the effects of applying carrier-selective contacts to various Si solar cells using Synopsys Technology Computer-Aided Design (TCAD). These simulations consider both intrinsic recombination (Auger and radiative recombination) and extrinsic recombination (Shockley-Read-Hall (SRH), surface and contact recombination). The simulated cell structure is illustrated in **Figure 1**a, which consists of a lightly p-doped bulk absorber with a doping concentration of $10^{16}$ cm$^{-3}$, and an n$^+$ emitter and a p$^+$ back surface field (BSF) layer, each with a doping concentration of $10^{20}$ cm$^{-3}$ and thickness of 50 nm. The cell thickness varies from 1 μm to 100 μm and the minority carrier lifetime is assumed to be 1 ms. In order to demonstrate the effects of carrier-selective contacts, the minority carrier recombination velocity at the contacts is set at $10^2$ cm-sec$^{-1}$ for both electron and hole selective contacts and $10^7$ cm-sec$^{-1}$ for metal contacts. The surface recombination velocity (SRV) at the passivated surfaces is assumed to be 10 cm-sec$^{-1}$. The recombination currents are analyzed under different mechanisms, including bulk, BSF, emitter, and contact recombination. The bulk recombination current accounts for the SRH, Auger and radiative recombination in the bulk region. The front/rear recombination current includes all the recombination at the emitter/BSF and at the front/rear interfaces. The contact recombination current includes the recombination at both front and rear contacts.

The $V_{oc}$ of cells with and without carrier-selective contacts are plotted versus the cell thickness in Figure 1b. It can be clearly observed from this plot that the carrier-selective contact can improve the $V_{oc}$ of cells with thickness from 1 μm to 100 μm, with the enhancement far more prominent in the thinner cells. In the 1 μm cell, the carrier-selective contact improves the $V_{oc}$ by 60 mV, while the carrier-selective contact improves the $V_{oc}$ by 12 mV in the 100 μm thick cell. Such a difference in $V_{oc}$ enhancement for different cell thicknesses can be explained by analyzing the components of the recombination current density $J_0$, as shown in Figure 1c. In a 1 μm thick cell with metal contacts, the contact recombination current ($J_{0,contact}$) is significantly larger than the other components, contributing to ~90% of $J_0$. Applying carrier-selective contacts effectively suppresses this $J_{0,contact}$ and reduces $J_0$ from 12 fA-cm$^{-2}$ to 1.2 fA-cm$^{-2}$, resulting in a $V_{oc}$ improvement of 60 mV in the 1 μm thick cell. In the 100 μm thick cell with metal contacts, bulk and contact recombination current each contribute to ~40% of the total recombination current. Carrier-selective contacts can therefore only reduce $J_0$ from 24 fA-cm$^{-2}$ to 15 fA-cm$^{-2}$ and improve $V_{oc}$ by 12 mV, which is less significant than the thin cells.

On the other hand, applying carrier-selective contacts enhances the increase in $V_{oc}$ achieved through decreasing the cell thickness. In the metal contact cell, the $V_{oc}$ is increased by only 20 mV as the cell thickness decreases from 100 µm to 1 µm, while that change is enhanced to 60 mV in the carrier-selective contact cells. Especially, the $J_{0,\ contact}$ in the metal contact cells becomes the majority term in $J_0$ when the cell is thinned down below 10 µm. Since such recombination does not scale with the change of thicknesses, the increase of $V_{oc}$ becomes saturated in the metal contact thin cells. Applying carrier-selective contacts can significantly reduce $J_{0,contact}$, leading to a regime where the bulk recombination becomes the dominant recombination mechanism. The benefit of $V_{oc}$ from the decrease in cell thickness remains in all cases, even in extremely thin cells. Therefore, carrier-selective contacts are very important for achieving high voltage and high efficiency in thin c-Si cells with thickness below 10 µm.

As shown in **Figure 2**a, an ultra-thin film solar cell is fabricated on a silicon-on-insulator (SOI) wafer. The buried oxide functions to block any carriers generated in the thick Si substrate. The active region of the cells is epitaxially deposited by chemical vapor deposition (CVD). The emitter is a layer of 100 nm thick $10^{19}$ cm$^{-3}$ phosphorus doped n$^+$ Si; the base is a layer of 1800 nm thick lightly (~$10^{16}$ cm$^{-3}$) boron doped Si; and the back side filed (BSF) is a layer of 100 nm thick $10^{19}$ cm$^{-3}$ boron doped p$^+$ Si. A 5 nm thick TiO$_2$ layer is deposited by Atomic Layer Deposition (ALD) between the emitter and the metal to form a top MIS contact. A metal contact is applied to the BSF layer as a bottom contact.

Figure 2c shows the current-density-voltage ($J$-$V$) characteristics of the devices measured under air-mass 1.5 solar spectrum. The 2 μm thick c-Si cell with a TiO$_2$ MIS contact achieves a $V_{oc}$ of 645mV, a $J_{sc}$ of 16.7 mA-cm$^{-2}$, and an efficiency $\eta$ of 8.9%. Compared to the cell with metal contacts, the cell with MIS contacts demonstrates an enhancement in both $V_{oc}$ and $J_{sc}$. The $V_{oc}$ of 645 mV in the MIS contact cell is 10 mV higher than that of the metal contact cell, which is mainly due to the reduced contact recombination loss. The quasi-steady-state $V_{oc}$ (QSS$V_{oc}$) characterization shows a recombination current density ($J_0$) of 0.46 pA-cm$^{-2}$ in the MIS contact cell and a $J_0$ of 0.72 pA-cm$^{-2}$ in the metal contact cell. This 40% drop of $J_0$ reflects a significant reduction in the recombination velocity at the contact interface under the TiO$_2$ MIS contacts. For a cell with a thickness of 2 μm, our simulation suggests that the $J_{0,contact}$ contributes to 80% - 90% of $J_0$. The electron-selective contact can suppress the recombination at the top contact and thus reduce the $J_{0,contact}$ by half, resulting in a ~40% reduction of $J_0$ and consequently the 10 mV $V_{oc}$ enhancement in our cell with TiO$_2$ MIS contacts.

The MIS contact cell shows a 0.9 mA-cm$^{-2}$ higher $J_{sc}$ than the metal contact cell, which is mainly a result of the improved carrier collection efficiency due to the suppressed minority carrier recombination at the TiO$_2$ MIS contact. Figure 2d shows the measured external quantum efficiency (EQE) of cells with TiO$_2$ MIS contacts and with metal contacts. It can be seen that the response of the MIS contact cell is clearly improved in the wavelength range of 400 nm – 550 nm. Since photons in this wavelength range are absorbed near the top surface where the TiO$_2$ MIS contact suppresses recombination, the carrier collection efficiency is thus improved, resulting in higher EQE and a $J_{sc}$ improvement of 0.9 mA-cm$^-$

$^2$ compared with the metal contact cell. In the long wavelength range, the EQE of the 2 μm thick cell is considerably decreased due to the weak absorption of Si. As suggested in previous research, this issue can be resolved by applying nano-scale light trapping structures [1-3,23-25].

The key to the $TiO_2$ MIS contact is the property of interface passivation at the $TiO_2$/n-Si interface, which can be improved by post deposition annealing [26]. Samples with $TiO_2$ MIS contacts were annealed in forming gas at different temperatures ranging from 300 °C to 500 °C for 90 sec. In **Figure 3**, it can be seen that as the annealing temperature increases from 300 °C to 450 °C, $J_{sc}$ increases significantly by 13% and $J_0$ drops by 43%, which is equivalent to 16 mV increases in $V_{oc}$. These results show that annealing improves the passivation effect of $TiO_2$ as well as the EQE at short wavelengths. As the annealing temperature further increases to 500 °C, $J_0$ increases by 38% and $J_{sc}$ drops by 3% compared to the optimized condition. These are probably due to degradation in the passivation effect, which is caused by a phase change of $TiO_2$ from anatase to rutile at high temperature [26, 27]. Annealing at 450 °C produces the best device with the lowest $J_0$ of 4.6 pA-cm$^{-2}$, highest $J_{sc}$ of 16.7 mA-cm$^{-2}$ and best efficiency of 8.9%. The performance of this device is detailed in the first row of Table 1.

In summary, we have demonstrated a carrier-selective contact with a $TiO_2$ MIS structure. With this carrier-selective contact, a 2 μm thick Si solar cell can achieve a $V_{oc}$ of 645 mV, which is 10 mV higher than that of a comparable cell with metal contact. We use TCAD simulations to analyze the recombination loss in thin cells and reveal the origin of the $V_{oc}$ improvement by the MIS contact. Our results show contact recombination is the major recombination mechanism in thin cells and eliminating this source is essential to the design of high-efficiency c-Si cells. We also studied the post deposition annealing of $TiO_2$ MIS contact cells and point out a method to improve the effect of this carrier-selective contact. Over all, this work demonstrates a new design for carrier-selective contacts and its application in ultra-thin-film c-Si solar cells, and provides a path to high-efficiency ultra-thin-film c-Si solar cells.

**Experimental Section**

Our ultra-thin-film c-Si solar cells are fabricated on a silicon-on-insulator (SOI) wafer to precisely control the cell thickness. The buried oxide blocks any carriers generated in the thick Si substrate. The active region of the cells is epitaxially deposited by chemical vapor deposition (CVD). First, a layer of 100 nm thick $10^{19}$ cm$^{-3}$ boron doped p$^+$ Si and a layer of 400 nm thick $10^{17}$ cm$^{-3}$ boron doped Si are grown to form the BSF. Then 1000 nm thick lightly (~$10^{16}$ cm$^{-3}$) boron doped base is grown. Last, a 400 nm thick $10^{18}$ cm$^{-3}$ phosphorus doped n$^+$ Si region and a 100 nm thick $10^{19}$ cm$^{-3}$ phosphorus doped n$^+$ Si region are grown to form the emitter. After this junction formation, trench isolation and mesa definition were done by photolithography and dry etch. The wafer is then oxidized in water vapor at 950ºC to grow 100nm thick thermal oxide to passivate the Si surface. The top contact pattern is then defined by lithography, and thermal oxide is wet etched to open the contact area. 5 nm thick $TiO_2$ is deposited by atomic layer deposition (ALD) using a precursor of Tetrakis (dimethylamido) titanium (IV) and steam at 200ºC. A metal contact of 10 nm thick titanium (Ti) and 200nm thick aluminum (Al) is e-beam evaporated immediately after the deposition of $TiO_2$ to prevent degradation. After metallization, reactive-ion etching (RIE) is performed to etch away all the remaining $TiO_2$ on the wafer to prevent shunting. The bottom contacts are defined by lithography and formed by the evaporation of Ti (20 nm) and Al (300 nm). To optimize for best performance, cells are annealed under forming gas at different temperatures in the range from 300°C to 500°C for 90 seconds. Finally, the thermal oxide is thinned down to ~80 nm by dry etch to function as the single layer anti-reflective coating. Control samples without the $TiO_2$ layer are fabricated with an identical process except that the $TiO_2$ deposition step is skipped.

Solar cell efficiency is measured under AM 1.5G normal illumination (1000 W/m$^2$, 1 sun) at room temperature. A standard solar simulator is used as the light source, with its intensity monitored by a certified photodetector. For the EQE measurement, a mechanically chopped monochromatic light beam is used as the light source, and the photocurrent is measured using a lock-in amplifier. The light intensity for the EQE measurement is calibrated with an amplified, calibrated photodetector. The saturation current is measured using the quasi-steady-state open circuit voltage (QSS$V_{oc}$) method with a Sinton Instrument WCT-120.

**Acknowledgment**

This work was supported by the Bay Area Photovoltaic Consortium (BAPVC) and the Global Climate and Energy Project (GCEP) at Stanford University. The authors acknowledge the Stanford Nanofabrication Facility (SNF) for the use of the processing facilities, P. Beck for experimental support, and J. Renshaw and S. Chatterjee for characterization support.

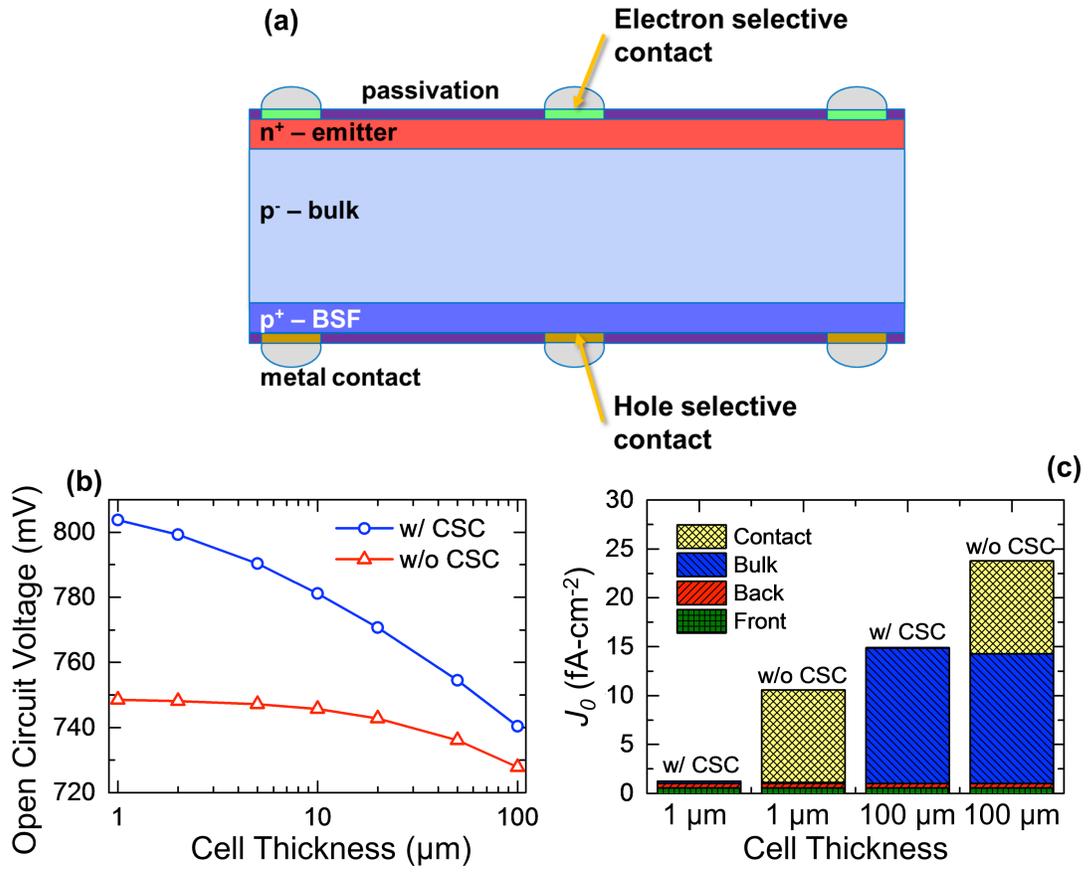

**Figure 1.** (a) Schematic illustration of the simulated c-Si solar cell with carrier selective contacts. (b) Simulation result of $V_{oc}$ as functions of the cell thickness ($W$) with (blue circle) and without (red triangle) carrier-selective contacts (CSC). (c) Recombination current density ($J_0$) breakdown for 1 μm and 100 μm thick Si solar cells with and without CSC.

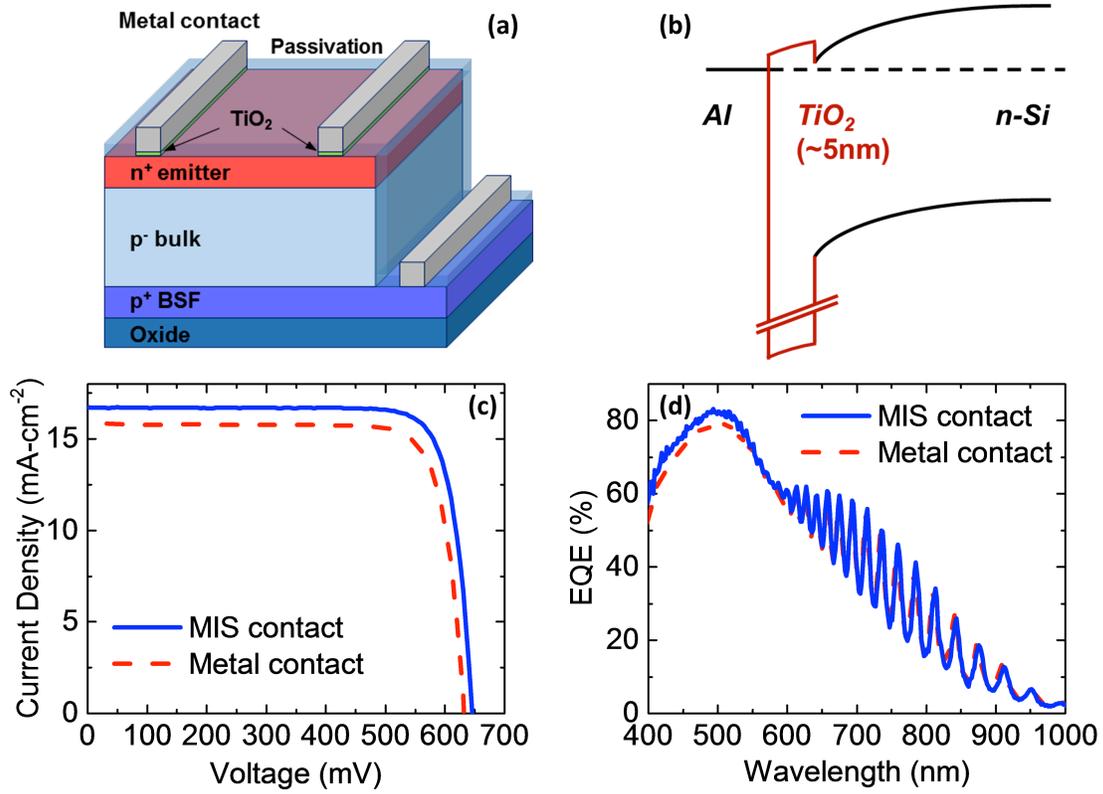

**Figure 2.** (a) Schematic of the ultra-thin-film c-Si solar cell with the $TiO_2$ MIS contact. (b) Band diagram of the $TiO_2$ MIS contact. (c) Current-density-voltage characteristics of the ultra-thin-film c-Si solar cells with $TiO_2$ MIS contacts (blue solid line) and with metal contacts (red dash line). (d) External quantum efficiency (EQE) of 2 μm thick Si solar cells with $TiO_2$ MIS contacts (blue solid line) and with metal contacts (red dash line).

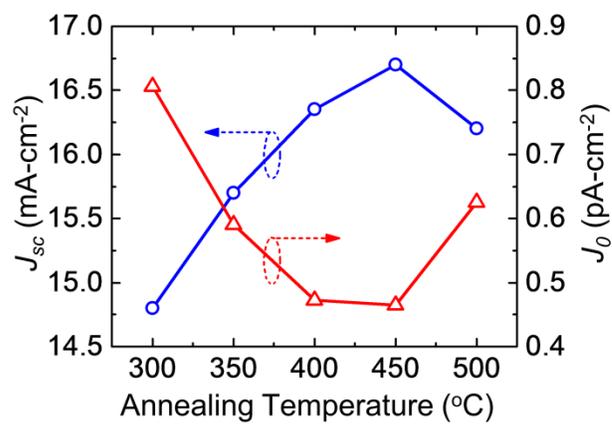

**Figure 3.** Measured $J_{sc}$ (blue circle) and $J_0$ (red triangle) as a function of annealing temperature.

**Table 1.** Key performance parameters of 2 μm thick Si solar cells

|  | $V_{oc}$ (mV) | $J_{sc}$ (mA-cm$^{-2}$) | F.F. (%) | $\eta$ (%) | $J_0$ (pA-cm$^{-2}$) |
|---|---|---|---|---|---|
| TiO$_2$ MIS Contact | 645 | 16.7 | 82.2 | 8.9 | 0.46 |
| Metal Contact | 635 | 15.8 | 81.6 | 8.2 | 0.72 |